\documentclass[preprint,pre]{revtex4-1}
 
\usepackage{graphicx}
\usepackage{amsmath}
 
\usepackage{amssymb}
\usepackage{natbib}
 
\usepackage{color}

\begin{document}

\newcommand{\inl}[1]{$ #1 $}

\title{The role of particle shape in active depletion}

\author{J. Harder$^{1}$}
\thanks{These three authors  equally contributed to this work}
\author{S. A. Mallory$^{1}$,$^*$ C. Tung$^{1}$ $^*$}

\author{C. Valeriani$^{2}$ and A. Cacciuto$^1$}
\email{ac2822@columbia.edu}
\affiliation{$^{1}$Department of Chemistry, Columbia University\\ 3000 Broadway, New York, NY 10027\\ }
\affiliation{$^{2}$Departamento de Quimica Fisica, Facultad de Ciencias Quimicas, Universidad Complutense de Madrid, 28040 Madrid, Spain}

\begin{abstract}
Using numerical simulations, we study how a solution of small active disks, acting as depletants, induces effective interactions on large passive colloids. Specifically, we analyze how the range, strength, and sign of these interactions are crucially dependent on the shape of the colloids. Our findings indicate that while colloidal rods experience a long-ranged predominantly attractive interaction, colloidal disks 
feel a repulsive force that is short-ranged in nature and grows in strength with the size ratio between the colloids and active depletants.  For colloidal rods, simple scaling arguments are proposed to characterize the strength of these induced interactions. 
\end{abstract}

%%\keywords{latex-community, revtex4, aps, papers}

\maketitle

{\it Introduction} -- Complex fluids and colloidal mixtures are some of the most ubiquitous substances on our planet.  Aerosols, foams, emulsions, and gels have countless applications, and are the subject of 
intense scientific research across all disciplines.  
Recently, self-propelled or active colloidal systems have garnered considerable interest because of their exciting rheological properties and unusual phenomenological behavior.  
In contrast to traditional non-equilibrium systems, where directional driving forces emerge as a result of global changes of the thermodynamic variables or boundary conditions such as temperature and pressure, active systems are intrinsically out of equilibrium at the single particle level. 
The combination of this unique non-equilibrium driving force and the inherent stochastic nature of  microscopic processes have endowed active systems with remarkable collective behavior.  Self-propulsion is typically achieved by conversion of chemical or ambient free energy into consistent, directed motion.   There are numerous examples of biological and synthetic active systems at the nanoscale, including the cytoskeleton of eukaryotic cells \cite{walter_2008}, bacterial suspensions, and catalytically activated colloidal particles \cite{yiping_2011, armand_2005, bocquet_2010, sen_2006}. 
In the latter case self-propulsion has been observed in platinum/gold and gold/nickel bi-laterally coated Janus nano-rods in the presence of $\mbox{H}_2\mbox{O}_2$ \cite{sen_2006, whitesides_2002, ozin_2005}, and it is also achieved in colloids where enzymatic reactions take place on one side of a particle \cite{heller_2005}, or can be driven by self-thermophoresis in defocused laser beams \cite{sano_2010}.

Although significant work has been carried out to understand the phenomenological behavior of self-propelled systems (for a recent review of the subject we refer the reader to reference \cite{simha_2013}), we still
have  a poor understanding of how immersion into an active environment can affect the dynamic self-assembly pathways of 
large non-active bodies. This is a very important question in colloidal science where 
effective interactions (i.e. solvent mediated interactions) play a crucial role in stabilizing or driving self-assembly of colloidal particles.

One of the simplest ways of inducing a short range attraction among colloids is by taking advantage of the depletion effect which is an effective interaction achieved by the addition of numerous small, non-adsorbing components such as polymers (colloid-polymer mixtures) or colloids (asymmetric binary mixtures). 
The strength of this attraction increases linearly with the depletants' concentration (the small particles) and the range is comparable to the depletants' diameter. This attractive force is purely entropic and is due to an osmotic pressure difference when depletants are expelled from the region between two colloids \cite{ao_1958}.  In the simplest case where ideal polymers are used as depletants, this attraction takes the general form $F(r)=  \pi \rho k_B T  R^2(1-(r/2R)^2)$, where $r$ is the center-to-center distance between two colloids, $R$ is the colloidal radius, $\rho$ the density of depletant, and $T$ is the system temperature.  
If $\sigma$ is the diameter of the depletant, then the force between two colloids is present as long as $r \leq(2R+\sigma)$. For sufficiently large attractions,  usually controlled by the depletant's concentration, phase separation will occur~\cite{marchetti2,catesx}. The overall phase behavior as a function of polymer size and concentration has been thoroughly studied within the Oosawa-Asakura approximation \cite{ao_1958,evans_1999_jpcm, evans_1999_pre, fortini_2006}.  More recently there has also been an effort to characterize this force when the system is no longer in equilibrium \cite{likos_2003}, and 
a few studies have considered the phase behavior of active particles in a system of passive depletants ~\cite{cacciutoCates,Egorov}. 

In our previous work~\cite{cacciuto_2014}, we studied the thermomechanical properties of an active gas, and found that the force acting on two rods kept at a constant separation in the presence of active depletants has an anomalous, non-monotonic dependence on the temperature $-$ a notable deviation from the typical behavior of equilibrium systems. 
Two recent studies~\cite{reichhardt_2014, Rojas, bolhuis_2014} further revealed that using active particles as depleting agents can give rise to behavior that is drastically different from that induced by passive depletants.
In these works the forces induced by an active bath on two plates of a given length were measured, and layering effects and mid-to-long range interactions between plates were reported to develop when increasing  the self-propulsion. Additionally, Angelani et al. \cite{dileonardo_2011} have recently shown that the depletion attraction alone cannot describe the effective interactions between passive colloids in a bath of active particles. 
In a way, it is therefore inaccurate to refer to these forces as {\it active depletion}, but we will nevertheless carry on with this nomenclature throughout the paper to keep the analogy with the parent equilibrium system.

In this paper, we go one step further and show how the strength, the sign and the range of this effective interaction can be controlled by tuning the geometry of the passive bodies in a way that is very different from what would be expected of their equilibrium counterparts. Specifically, we characterize how the effective interaction between two colloidal particles varies as a function of the magnitude 
of the self-propelling force of the depletant and the depletant-to-colloid size ratios.
In addition, we highlight the strikingly different nature of the induced interaction 
when the colloids consist of rods or disks.
\newline

{\it Model} -- We consider a two dimensional system of large, passive, colloidal particles of diameter $\sigma_c$ immersed in a bath of smaller active particles of diameter $\sigma$ and unit mass $m$ at a volume fraction $\phi_b$.  Each active particle undergoes Langevin dynamics at a constant temperature, $T$, while self-propulsion is introduced through a directional force which has a constant magnitude $F_a$, along a predefined orientation vector, $\boldsymbol{n}=[\sin(\theta), \cos(\theta)]$.  The equations of motion of a bath particle are given by the coupled Langevin equations

\begin{equation}
m \ddot{\boldsymbol{r}} =-\gamma  \dot{\boldsymbol{r}}-\partial_{\boldsymbol{r}} V(r)+\sqrt{2\gamma^2D} \xi(t)+F_a\boldsymbol{n}  \hspace{1 cm} \mbox{and}  \hspace{1 cm}
\dot{\theta}  = \sqrt{2D_r} \xi_r(t)
\end{equation}
 
\noindent where $\gamma$ is the friction coefficient,  $V$ the total conservative potential acting between any pair of particles,  $D$ and $D_r$ are the translational and rotational diffusion constants, respectively (with $D_r=3D/\sigma^2$).  The typical solvent induced Gaussian white noise terms for both the translational and rotational motion are characterized by $\langle  {\xi}_i(t) \rangle = 0$ and   $\langle  {\xi}_i(t) \cdot   {\xi}_j(t') \rangle = \delta_{ij}\delta(t-t')$ and $\langle  {\xi_r}(t) \rangle = 0$ and   $\langle  {\xi_r}(t) \cdot   {\xi_r}(t') \rangle = \delta(t-t')$, respectively.

Bath particles are disks with diameter $\sigma$ which interact with each other via the Weeks Chandler Andersen (WCA) potential
\begin{equation}
 V(r_{ij})=4 \epsilon \left[ \left( \frac{\sigma_{ij}}{r_{ij}} \right)^{12}- \left( \frac{\sigma_{ij}}{r_{ij}} \right)^{6}+\frac{1}{4} \right]  
  \label{eq:LJ_V}
\end{equation} 
\noindent with a range of interaction  extending out to $r_{ij}=2^{1/6}\sigma$. Here $r_{ij}$ is the center to center distance between any two particles $i$ and $j$, and $\epsilon$ is their interaction energy. 

\begin{figure}[htbp]
\centering
\includegraphics[scale=1]{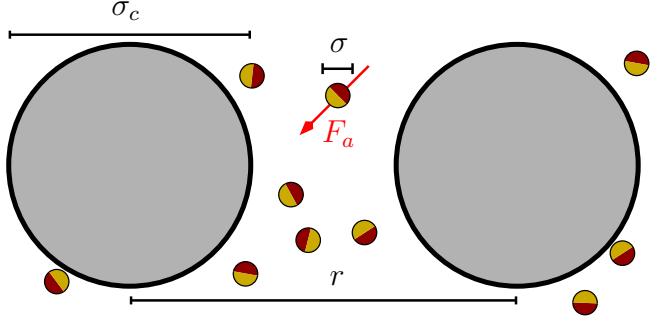}
\caption{Schematic representation of two colloidal disks in a bath of active particles.  The smaller, active components move according to Eq.1. The  persistent force  $F_a$  acts along a defined axis (as shown by the arrow as well as the colors, where red corresponds to the back of the particle and yellow the front).}\label{F1}
\end{figure}

Suspended colloids are either rods or disks.  The large colloidal disks interact with the bath particles through the same WCA potential defined above, with $\sigma_{ij}=(\sigma+\sigma_c)/2$, where $\sigma_c$ is the colloidal diameter. The rods  are modeled as rectangular regions of width $\sigma_w=2.5\sigma$ and vertical length $\ell$, and also repel the particles according to Eq.~\ref{eq:LJ_V}, where the separation $r_{ij}$ is the smallest distance between the particle the wall. Figure ~\ref{F1} shows a sketch of the model for disks.
The strength of interaction for both the depletant-depletant interaction and the depletant-colloid interaction is chosen to be 
$\varepsilon=10\,k_{\rm B}T$. The simulation box is a square with periodic boundary conditions, the Langevin damping parameter is set to $\gamma=10\tau_0^{-1}$(here $\tau_0$ is the dimensionless time), and the timestep to $\Delta t=10^{-3}\tau_0$. 
Each simulation is run for a minimum of $3 \times10^7$ iterations.  All simulations were carried out using the numerical package LAMMPS~\cite{plimpton_1995}, and 
 throughout this work we  use the default dimensionless Lennard Jones units as defined in LAMMPS, for which the fundamental quantities mass \(m_0\), length \(\sigma_0\), epsilon \(\epsilon_0\), and the Boltzmann constant \(k_{B}\) are set to 1, and all of the specified masses, distances, and energies are multiples of these fundamental values. In our simulations we have $T=T_0 = \epsilon_0/k_{B}$, $m=m_0$, $\sigma=\sigma_0$, and  $\tau_0 = \sqrt{\frac{m_0 \sigma_0^2}{\epsilon_0}}$.

% \newline

{\it Results} -- 
We first present the results for disk-shaped colloids.
To understand the effective interactions induced by active bath particles on the suspended colloids, we proceed in two ways:
(1) We measure the radial distribution function $g(r)$ for a suspension of passive colloidal disks in the presence of the active depletants. In this case the large colloids move according to the Langevin dynamics in Eq.1, 
but with $\beta F_a\sigma=0$  $(\beta\equiv(k_{\rm B}T)^{-1}$), and without the rotational component.
(2) We calculate the effective force between two  colloids  by directly measuring the mean force acting on the particles when they are frozen in place
as a function of the bath  activity, colloidal shape (disks and rods), and  colloid separation $r$.

\begin{figure}[h!]
\centering
\includegraphics[scale=1]{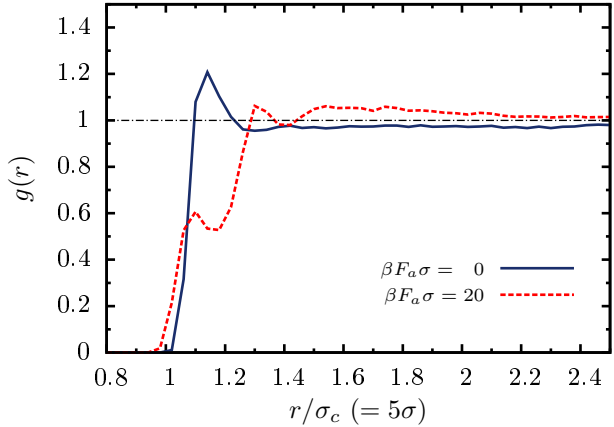}
\caption{Radial distribution function $g(r)$ of the large colloids for two different values of self-propulsion $\beta F_a\sigma$.  In a passive bath (blue, solid) the expected peak signifies a short ranged attraction between colloids.  In an active bath (dashed line), $g(r)$ takes values which are less than $1$, suggesting a repulsion between the colloids.  This repulsion increases with the bath's activity.}
\label{fig:rdf}
\end{figure}

For non-active equilibrium systems, the reversible work theorem provides a simple relationship between the potential of mean force and the radial distribution function, namely $U(r)=-k_{\rm B} T \log[g(r)]$ \cite{chandler_book}.
Unfortunately, such a relation does not necessarily hold in the presence of an inherently out-of-equilibrium active bath.  Nevertheless, from the $g(r)$ it is possible to extract qualitative information about the sign, strength and range of the interaction.
To determine the $g(r)$, simulations were carried out with 100 colloidal disks  of diameter $\sigma_c=5\sigma$ immersed in an active bath at a volume fraction $\phi_b=0.1$, and the simulation box is a periodic cube with box length 150$\sigma$. The resulting radial distribution functions are shown in Fig. \ref{fig:rdf}.  Each simulation was run for over $10^8$ time-steps.

In the passive system with $\beta F_a\sigma = 0$, the $g(r)$ presents a large peak at the colloid contact separation as expected for this system which is characterized by a strong depletion attraction.  In other words, this peak indicates a significant likelihood of finding two colloids in contact with each other.  When the bath is active, however, the radial distribution function is smaller than $1$ for small colloid separations, which strongly suggests that there is an effective repulsion between the colloids.  

\begin{figure}[h!]
\centering
\includegraphics{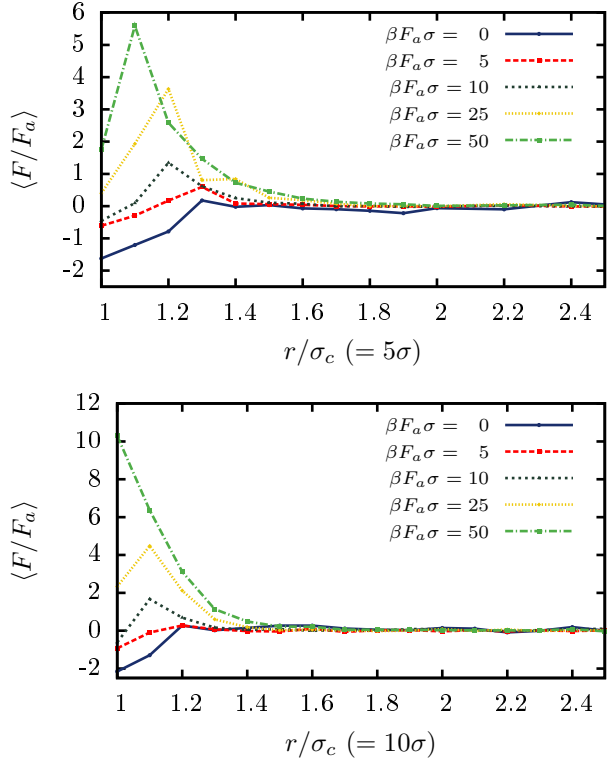}
\caption{Effective rescaled forces $ \langle F/F_a\rangle$ experienced by two colloidal disks as a function of separation 
for different values of depletant's activity. (a) Shows the result for $\sigma_c=5\sigma$ and (b) for $\sigma_c=10\sigma$.
For both sets of simulations $\phi_b = 0.1$. Rescaling is only applied as long as $\beta F_a\sigma\neq 0$.  Positive values correspond to a repulsion, which clearly dominates any depletion driven interaction when the bath is active.  The larger the active force and the larger the colloid-to-depletant size ratio is, the stronger the repulsion.}
\label{fig:disk_force}
\end{figure}

To provide a more quantitative measurement of this repulsion and to better understand its nature, we proceed by performing simulations where two colloids are frozen in place and the force between them is measured directly from their interactions with the active bath particles. All results presented below are obtained at a constant volume fraction  $\phi_b=0.1$.
The net force exerted on the two disks by the bath along the inter-colloidal axis was evaluated for two different colloidal sizes $\sigma_c=5\sigma$ and $\sigma_c=10\sigma$. The results are shown in Fig. \ref{fig:disk_force}.  

\begin{figure}[htbp]
\centering
\includegraphics[scale=1]{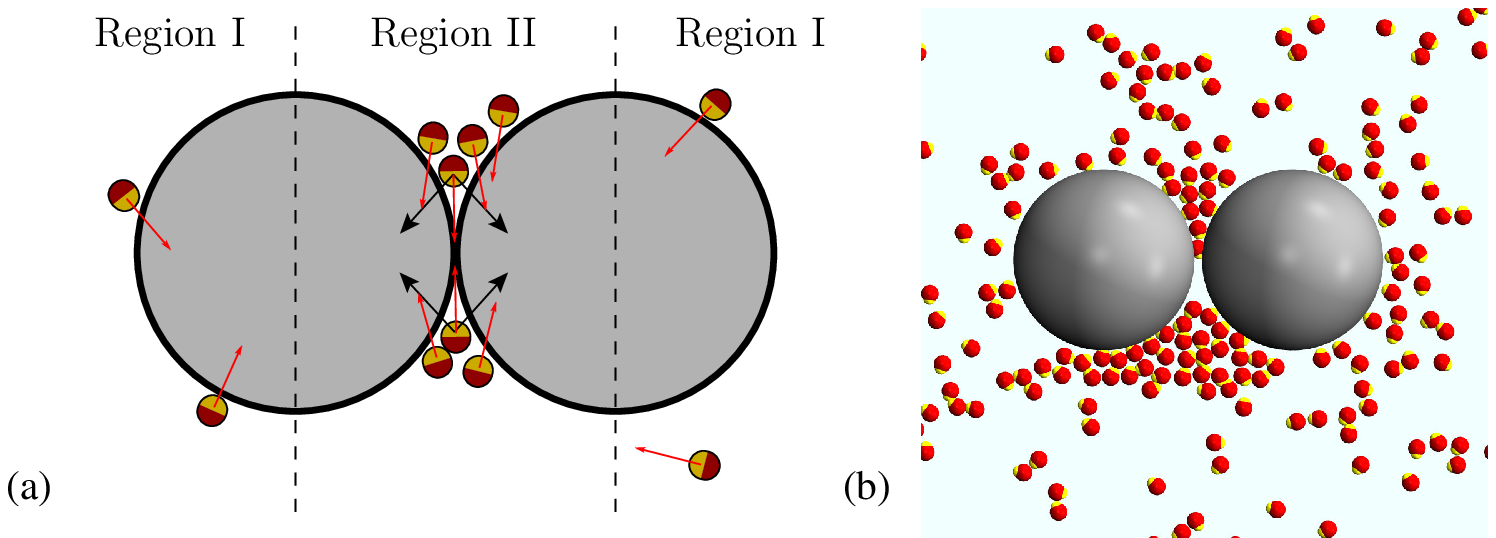}
\caption{(a) Sketch showing the effective forces exerted on the disks by the active particles in the two different regions. (b) Snapshot from a simulation of two colloidal disks of diameter $\sigma_c = 10\sigma$, at $\beta F_a\sigma =  50$.  The two large disks experience a net repulsion due to the trapping of active bath particles.  As in (a), the yellow portions of the particles indicate the direction along which the propulsive force is applied.\cite{VMD}}
\label{fig:disk_regions}
\end{figure}
In a passive bath, the interaction between two large colloids is well understood and is given by the depletion attraction previously discussed.  Surprisingly, as the bath becomes increasingly active, the effective interaction between the colloids becomes purely repulsive.  This result is consistent with the observed behavior of the $g(r)$.
The introduction of activity results in a repulsive force much larger than the depletion attraction observed in passive systems, and grows  with the extent of the activity. Notice however, that the range of the interaction is rather insensitive to the propulsion strength, and extends to a distance of roughly half the colloidal diameter. 

To better understand this phenomenon, we examine the duration of collisions between bath particles and the colloidal disks as well as where along the colloids' surface these collisions take place.  
Here, we define the inner surface of a colloid (Region II Fig. \ref{fig:disk_regions}(a)) as the half circle which lies closer to the center of the other colloid, and the outer surface (Region I Fig. \ref{fig:disk_regions}(a)) as the half circle which is further away from the other colloid.
 When a bath particle strikes the outer surface of either of the large disks a force is generated with a net component, $F_{I}$, which pushes the disks toward each other.  When a particle strikes the inner surface of either disk it generates a force with components, $F_{II}$, which pushes the disks away from each other (see Fig. \ref{fig:disk_regions}(a) for a sketch of these forces).
 The effective force experienced by the two disks is determined by the number of particles at the surface of each region as well as by the average duration of a collision event.  

Unlike equilibrium systems for which one expects a particle to bounce off a wall upon collision, the collision of an active particle with a wall is similar to that of a car driving into a wall. The active particle will continue to exert a force into a barrier until its propulsion axis begins to rotate, upon which the particle will slide along the wall.  For a given strength of self-propulsion, the duration of a collision is controlled by the  rotational diffusion, which is governed by thermal fluctuations, and has a strong dependence on the local environment{}. 
During collisions, active particles remain in contact with the surface of the larger colloids for some amount of time before sliding off or rotating away.  The duration of contact is in large part determined by the geometry of the colloids.

When the colloids are far apart, active particles have equal probability of striking either their inner or outer surfaces, leading to a zero net effective force between them. 
When the colloids are in contact, they form an object characterized by regions of both positive and negative curvature.
The outer surfaces  have positive curvature, and colliding particles can slide off rather quickly.  
The inner surfaces have negative curvature and can create a trap~\cite{chantalmer} 
for the active particles, greatly increasing the duration of a collision.
The result is a net gradient  in particle concentration along the colloidal surface, leading to the effective repulsion reported in our simulations (See 
Fig.~\ref{fig:disk_regions}(b) for a snapshot from our numerical simulations).

\begin{figure}[h!]
\centering
\includegraphics[scale=1.0]{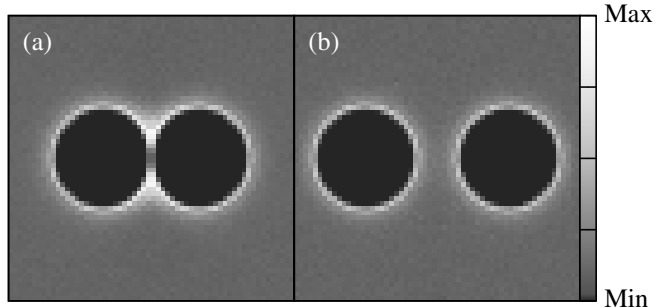}
\caption{Time averaged density maps of active particles for $\beta F_a\sigma = 50, \phi_b = 0.1$ around two colloidal disks of diameter $\sigma_c=10\sigma$ at a center-to-center distance of  $10\sigma$ (a), and $15\sigma$ (b).  In both cases, it is clear that active particles aggregate on the colloid surfaces.  When the colloids are in contact, a region of high active particle density can be seen near the effectively concave surface between the colloids.  The scale bar corresponds to the density of the bath particles, and goes from a minimum of zero in the space occupied by the colloids to a maximum value near the point where the colloids meet in (a).}
\label{fig:dmap_disk}
\end{figure}
To determine the surface concentration gradient of bath particles, we compute a density map of the active particles around the disks at large and small separations (Fig. \ref{fig:dmap_disk}).
As expected, when the disks are sufficiently far apart, there is no significant difference between the particle density  on the inner and outer region.  When the surface-to-surface separation between the disks is of the order of $2\sigma$ however, the inner density is significantly larger than the outer one resulting in the observed repulsion.

\begin{figure}[!h]
\centering
  \includegraphics[scale = 1]{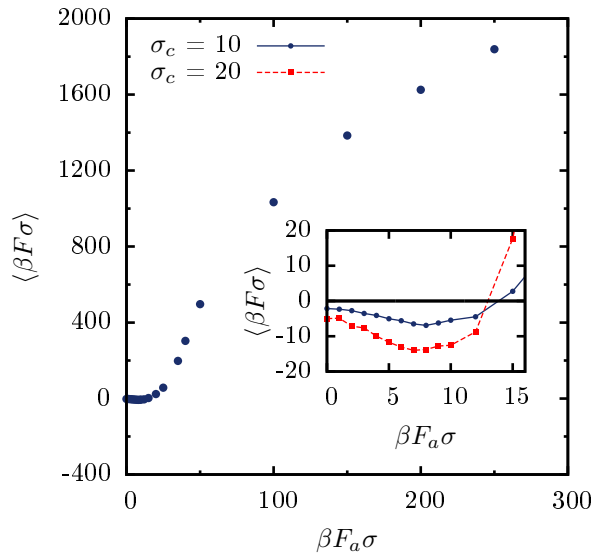}
  \caption{Effective force between two disks in contact and in the presence of active depletants as a function of self-propulsion $\beta F_a\sigma$ for disks of diameter $\sigma_c = 10\sigma$ and $20\sigma$, at  $\phi_b=0.1$ and  box side length $L=150\sigma$.  The inset shows that for moderate active forces, an enhanced attraction is observed. }
  \label{fig:disk_contact}
\end{figure}

We also measure the net force between two disks at contact as a function of the active force for 
$\sigma_c=10\sigma$ and $\sigma_c = 20\sigma$, this is shown in Fig. \ref{fig:disk_contact}.  In the absence of activity, we recover the expected depletion attraction for equilibrium systems, and in the limit of large active forces, we observe the repulsive behavior discussed above. However, for intermediate values of $\beta F_a\sigma \in [1,10]$, we observe a strengthening of the attraction between the colloids as a function of $F_a$. These results suggest that as long as $F_a$ is sufficiently small, the main effect of the propelling force is that of an effective higher temperature of the bath, leading to a strengthening of the depletion interactions. 
It should be noted that, for this range of active forces, the persistence length of the trajectory traced by a single active particle, estimated as $d\simeq (F_a/\gamma)D_r^{-1}=(\beta F_a\sigma^2)/3$, is significantly smaller than the colloidal diameter  used in these simulations. One possible interpretation of this result is that as long as $d\ll\sigma_c$ the colloids 
always experience an attractive interaction. To see whether this is true, we repeated our simulations for the same range of active forces for two larger colloids with twice the diameter $\sigma_c=20\sigma$, placed in contact with each other.
%We find that the contact force is larger for $\sigma_c=20\sigma$ than for  $\sigma_c=10\sigma$, and this is simply due to the corresponding increase of the size of the excluded area.
Surprisingly, the sign of the interaction switches over at approximately the same value of $F_a$ as for the smaller colloids (see Fig. \ref{fig:disk_contact} inset).  If the $d\ll\sigma_c$ argument were correct, the attraction should persist to larger values of $F_a$ for the larger colloids, but we find that this is not the case.
One reason for this could be due to the fact that larger colloidal diameters also correspond to larger regions where particles can be trapped.  This leads to an enhanced repulsion between the colloids that competes with the strengthened attraction. This enhanced repulsion is easily seen when comparing the two plots in Fig. \ref{fig:disk_force} showing that larger colloids experience overall larger repulsive forces.
Finally, it should  be noticed that the range of the interaction between disks is not very sensitive to the strength of the propelling force, and does not extend to separations much further than a fraction of the colloidal diameter.

We now turn our attention to the case of two colloidal rods.  A system composed of two such rods was one of the earliest to be studied in the context of the depletion attraction.  As was the case for disks, two rods in a bath of smaller particles experience an entropic attractive force which depends on the size of the excluded area, as well as the size and density of the depletants, and the temperature.  This force can be large when compared to that between two suspended disks due to the relatively larger excluded area when rods are in contact.  Unlike colloidal disks, rods have no curved surfaces, so active bath particles which come into contact with the surface of a rod are effectively confined to move along this surface until they rotate away or slide to the end of the rod.

\begin{figure}[!h]
\centering 
\includegraphics[scale = 1]{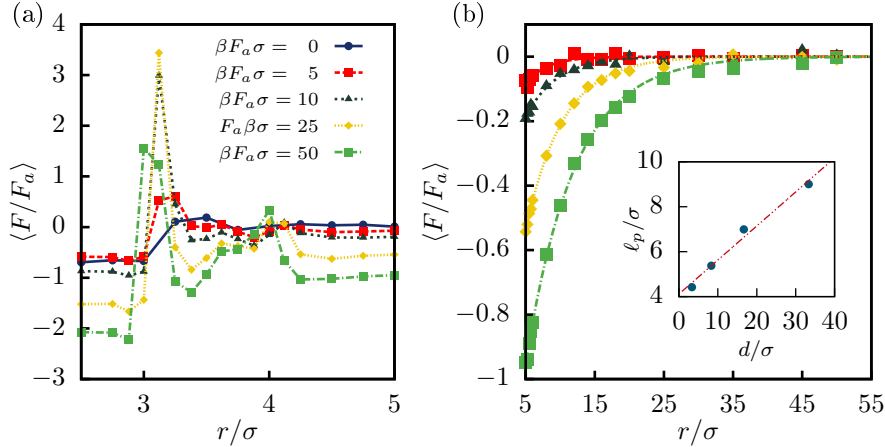}
\caption{Measured force (scaled by the active force) felt by two rods in a bath of active particles as a function of their separation for different strengths of self-propulsion. (a) Shows the behavior for small separations,
($r\sim \sigma$) while (b) shows the curve for large separations ($r\gg \sigma$).
The inset in (b) shows how the decay length $\ell_p$ of the fitted exponential curves in the limit of large separations (a measure of the interaction range)  grows linearly with the particle persistence length $d/\sigma=\beta F_a\sigma/3$.}
\label{fig:plates_force}
\end{figure}

When the separation between the rods is small, we observe an oscillating attractive and repulsive 
force. As also reported in \cite{bolhuis_2014}, this behavior is due to a competition between the forces exerted by the active particles on the outer surfaces, and the buildup of ordered layer of particles between the rods.
(see Fig. \ref{fig:plates_force}(a)).  
Surprisingly, at larger separations (Fig.~\ref{fig:plates_force}(b)), a large long-ranged attraction is induced between the rods. In agreement with~\cite{bolhuis_2014} 
this  attractive effective interaction can be well fit to an exponential and  
the range of the interaction is controlled by the effective persistence length of the path traced by the active particles 
$d/\sigma=(\beta F_a\sigma)/3$. The inset of  Fig.~\ref{fig:plates_force}(b) shows the linear dependence of the interaction decay length $\ell_p$ as a function of the particle persistence length $d/\sigma=\beta F_a\sigma/3$.  

\begin{figure}[!h]

\centering
  \includegraphics[scale = 1]{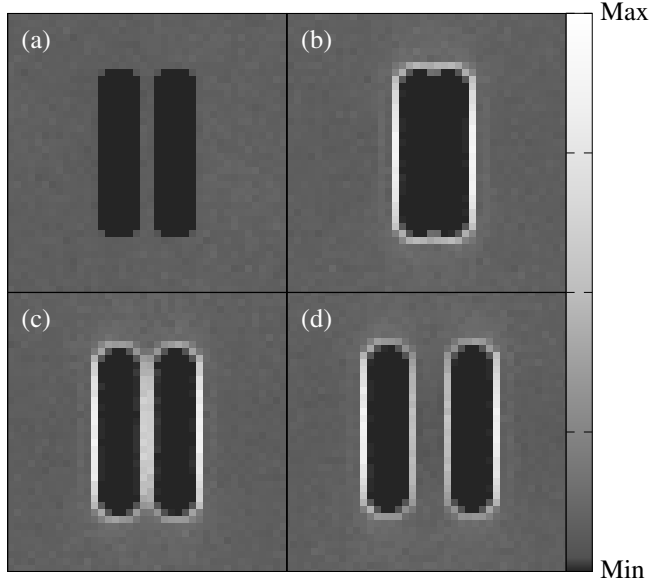}
  \caption{Time averaged density maps of active particles around larger rods. (a) $\beta F_a\sigma = 0$ at rod separation $r = 4\sigma$, (b) $\beta F_a\sigma = 50$ at rod separation $r = 2\sigma$, (c) $\beta F_a\sigma = 50$ at rod separation $r = 4\sigma$, and (d) $\beta F_a\sigma = 50$ at rod separation $r = 6\sigma$.  In the passive case (a), the bath particle density is  uniform across the simulation box, resulting in no long range interaction between the rods.  (b) shows the bath particle density profile around two rods in contact.  (c) shows  particles getting trapped in between the rods and giving rise to a net repulsion.  In (d), the density of bath particles on the outside of the rods is greater than that inside, leading to the observed long-ranged attraction.  As before, the scale bar shows the particle density, and goes from a minimum in the regions excluded by the rods to a maximum at points near the rod surfaces.}
\label{fig:dmap_rod}
\end{figure}

Further insight can be obtained by looking at the time-averaged density map of the depletants for different rod separations
 (Fig.~\ref{fig:dmap_rod}.)  
In the passive case, the density of bath particles is uniform throughout the simulation box.  However, when the bath particles are active, they aggregate on the surfaces of the rods and there is a marked difference in local density on the different rod surfaces.  Specifically, once the rods are separated by more than $\sim 4\sigma$, there are more bath particles on the outer surfaces than on the inner ones, resulting in an attractive force between the rods.  When the rods are at a smaller separation, the situation is the opposite{}.

In principle, this “shadowing” effect leading to a long-range attraction extending up to $\ell_p$ should also be observed for colloidal disks. However, in this case, partial layers of active particles form on the perimeter of the disks and can easily diffuse around it. This leads to a uniform density of active particles along the disk perimeter, which balances the forces acting on the colloids’ inner and outer sides. In simulations with ideal active particles, where no layering can develop, long-range attractive forces are observed for colloidal disks. This suggests that an equilibrium distance in the effective colloidal interactions will develop for sufficiently low concentrations of active particles.

An estimate of how the force exerted on two rods at contact scales with their length and with the strength of  the bath activity can be obtained with the following simple argument. In the diffusive limit (i.e. when the length of the rods is sufficiently large such that the particles can diffuse over their surface before sliding off) Fily et.~al \cite{hagan_2014}, have shown that for large self-propelling forces the typical time $t_1$  a particle remains in contact with a rod  scales as $t_1\sim (\frac{\gamma \ell}{ F_a})^{\frac{2}{3}} (\frac{1}{D_r})^{\frac{1}{3}}$.
 During this time the particle will exert an average force on the rod, that to leading order scales like $ F_a$. The time required for a particle  in a container of lateral size $L$ to find the rods can be estimated as $t_2\sim\frac{L^2}{\ell}\frac{\gamma}{ F_a}$, which accounts for the probability of finding the rod when moving at a speed $v_a=F_a/\gamma$ across the box. Alternatively, one can think of $1/t_2$ as the average collision rate between an active particle and the rod. So that 
$1/t_2=(1/L^2)Cv_a$, where $C=\ell$ is the cross section of the rod.
 
 During this time the particle will exert no force on the rod. The net average force can then be estimated as $\langle F \rangle\simeq \frac{ N F_a t_1}{t_1+t_2}$, ($N$ is the number of active particles) and for sufficiently large systems, $t_2\gg t_1$, it can be simplified to $\langle F \rangle \simeq N F_a t_1/t_2$, leading to the scaling behavior 
\begin{equation}
\langle F \rangle \simeq \rho \left(\frac{ F_a^{4} \ell^{5}}{\gamma D_r}\right )^{\frac{1}{3}}\label{scaling}
\end{equation} 
where \(\rho\) is the number density.  In the non-diffusive limit, when the rods are short and the force is so large that an active particle slides off of the surface before any diffusion can occur, $t_1\sim  \ell \gamma/ F_a$, the average force should scale as
\begin{equation}
\langle F \rangle \simeq \rho  F_a  \ell^{2}\label{scaling2}
\end{equation}  
For long rods or weak propelling forces, the residence time of the particles on the surface is simply 
controlled by the rotational diffusion $t_1=1/D_r$. In fact, in these cases a particle leaves the surface as soon as its axis turns away from the surface's normal, with $\langle \theta^2\rangle=(\pi/2)^2=2D_r t^{\rm max}_1$ as the upper bound. In this limit the average force should scale as 
\begin{equation}
\langle F \rangle \simeq \rho \frac{ F_a^2  \ell}{\gamma D_r}\label{scaling3}
\end{equation}  
Finally, whenever $t_1\gg t_2$ (for sufficiently high densities) one should expect to first order  
$\langle F \rangle \simeq  F_a (1-\frac{t_2}{t_1})$.

\begin{figure}[!h]

\centering
  \includegraphics[scale = 1]{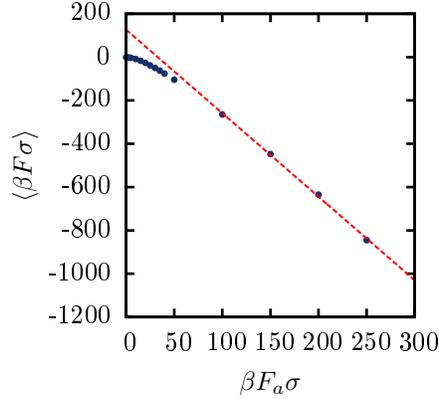}
  \caption{Effective force between two rods in contact with each other and in the presence of 
  active depletants as a function of self-propulsion $\beta F_a\sigma$ for rods of length $\ell=10\sigma$, at
  $\phi_b=0.1$ and  box side length $L=77\sigma$.  The dashed line is a linear fit to the net force at high bath activity and shows that our simulation results are consistent with Eq. \ref{scaling2}.}
  \label{fig:force_kick}
\end{figure}

\begin{figure}[!h]

\centering
  \includegraphics[scale = 1]{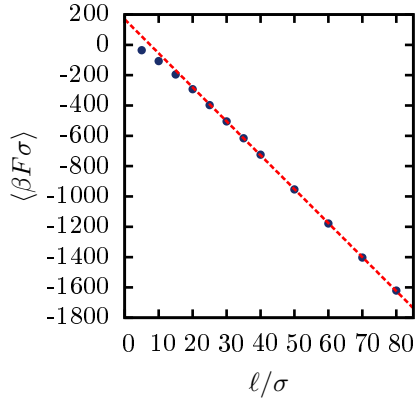}
  \caption{Effective force between two rods in contact with each other and in the presence of 
  active depletants as a function of rod length for  $\beta F_a\sigma = 50$, at $\phi_b=0.1$ and box side length $L=200\sigma$. A linear fit is also plotted, showing that our simulation results are consistent with Eq. \ref{scaling3}.}
  \label{fig:force_length}
\end{figure}

Deviations from this simple scaling are also expected at moderate and large densities due to the excluded volume interactions between particles. Figures~\ref{fig:force_kick} and~\ref{fig:force_length} show how the force between the rods scales with the strength of the activity and with the length of the rods in our simulations.  
Our numerical data have been taken at volume fraction $\phi_b=0.1$, which is sufficiently 
low to prevent any  bulk phase separation or aggregation of the active particles, yet, it is large 
enough to give non-negligible excluded volume effects. The relatively short length of the rods, $\ell=10\sigma$,
in these simulations implies that Eq.~\ref{scaling2} should give the most appropriate
description for the effective force.
This is consistent with Fig.~\ref{fig:force_kick}, that shows a linear dependence of the force  
with $ F_a$ in the large propulsion limit. When $ F_a$ is small, we expect $\langle F\rangle$ to
depend quadratically on $ F_a$. In fact, in this case the rotational diffusion is fast enough
to limit the  persistence-time of the particles on the colloidal surface, thus $ F_a$ will contribute  to the
average force  a quadratic (thermal-like)  term corresponding to an enhanced velocity of the particles~\cite{cacciuto_2014}.  

The analysis for the dependence on the length of the rods is a bit more complicated.
The problem is that the short $\ell$ limit is characterized by a short residence time 
that is inversely proportional to the self-propulsion, it grows linearly with $\ell$, and for which
clearly $t_2\gg t_1$. As $\ell$ becomes larger (up to $\ell=80\sigma$), $t_1$ should scale as $t_1\sim (\frac{\gamma \ell}{ F_a})^{\frac{2}{3}} (\frac{1}{D_r})^{\frac{1}{3}}$ (as long as  $t_1<1/D_r$), however, as the particles' residence time becomes longer, the average number of particles at contact becomes larger, and for moderate volume fractions, the assumption that  $t_2\gg t_1$ becomes less adequate.
To complicate matters even further, for large $\ell$, excluded volume interactions begin to matter, 
and local self-trapping of the particles may effectively increase of $t_1$ to values larger than $1/D_r$, so that the 
time to rotationally diffuse away from the surface may become faster than the time required to slide off the surface edge; 
making Eq.~\ref{scaling3} more appropriate in this regime.
This phenomenon is quite visible in our simulations for the longest rods ($\ell=80\sigma$), 
where  diffusive correlated motion of linear clusters of active particles over the rod surfaces takes place. 
At significantly lower densities than the ones considered in this paper,
we would have expected Eq.~\ref{scaling} to hold, but at $\phi_b=0.1$, all these effects become relevant.
We therefore expect a  combination of Eq.~\ref{scaling2} and Eq.~\ref{scaling3} to
provide a good approximation to our data.  Indeed, in the long 
rod limit the average force seems to be well fitted by a linear dependence on $\ell$.
\newline
 
{\it Conclusions} --
In this paper, we have studied the effective interactions induced by small active components on large passive colloidal particles as a function of the strength of the propelling force of the active bath
and of the geometry of the colloids. Our results indicate that the induced colloidal interactions 
are crucially dependent on their  shape, and that while a long ranged, predominantly attractive interaction is induced between rods, disks undergo a purely short range repulsion that grows in strength with the size ratio between the colloid and the active component. Crucial to this difference is the role of curvature, which 
determines whether  passive bodies act as traps or as efficient scatterers of active particles.
For instance, we have recently shown how curvature can be exploited to activate C-shaped passive bodies,
by creating density gradients across the colloids~\cite{cacciu_2014b}. 

Finally, as discussed above, we expect these interactions to be quite sensitive to the concentration of the active particles. Furthermore, we anticipate that many body effects play an important role in these systems, as the conditions leading to long-range attractive forces are strongly dependent on the specific
arrangements of colloids.

Although our study has been performed in two dimensions, the essence of
our results should be easily extendable to three dimensions when considering colloidal rods and spheres.
Our work further highlights the many differences between the effective
forces induced by small active components and those produced by the corresponding equilibrium system, and suggests that active depletion can have dramatic consequences on both
the phase behavior and the self-assembly of differently shaped colloids, with possible
applications in material engineering and particle sorting.
\newline

{\it Acknowledgments} --
 AC and CT acknowledge financial support from the National Science Foundation under Grant DMR-1408259. CV acknowledges financial support from a Juan de la Cierva Fellowship, from the Marie Curie Integration Grant 322326-COSAAC-FP7-PEOPLE-CIG-2012, and from the National Project FIS2013-43209-P. SAM acknowledges financial support from the National Science Foundation Graduate Research Fellowship (Grant No. DGE-07-07425). This work used the Extreme Science and Engineering Discovery Environment (XSEDE), which is supported by National Science Foundation grant number ACI-1053575.

%\bibliographystyle{apsrev4-1}
%\bibliography{act_dep_bib_final.bib}
%

\end{document}